\documentclass[twocolumn,showpacs,showkeys,aps,prl,superscriptaddress,nofootinbib]{revtex4}
\usepackage{graphicx}
\usepackage{amsmath}
\usepackage{epsfig}
\usepackage{lineno}
\usepackage{natbib}

\DeclareGraphicsExtensions{.epsi,.eps,.ps,.eps.gz,.ps.gz,.gif,.epsi.gz}
\newcommand{\su}     [1]  {\ensuremath{SU(#1)}}

\def\CPV {\ensuremath{CPV}}

\def\dz        {\ensuremath{\Delta z}}

\def\rhoz      {\ensuremath{\rho^{0}}}

\def\piz {\ensuremath{\pi^0}}

\def\kaon     {\tt Kaon}

\def\bteff{\beta_{\rm c,eff}}
\def\su3f{\ensuremath{SU(3)_{\rm\tiny flav}}}

\input superbsym

\long\def\inst#1{\par\nobreak\kern 4pt\nobreak
    {\it #1}\par\vskip 10pt plus 3pt minus 3pt}

\def\bea{\begin{eqnarray*}}
\def\eea{\end{eqnarray*}}
\def\sst{\scriptscriptstyle}

\def\fl{\ensuremath{f_{\sst L}}\xspace}

\begin{document}


{\pagestyle{empty}

\par\vskip 3cm

\title{
\Large \boldmath
Bounding hadronic uncertainties in $c\to u$ decays
}

\author{A. J. Bevan}
\affiliation{Queen Mary, University of London, Mile End Road, E1 4NS, United Kingdom}

\author{B. Meadows}
\affiliation{University of Cincinnati, Cincinnati, Ohio 45221, USA}

\date{\today}

\begin{abstract}
%
%
Time-dependent \CP asymmetry measurements in $D\to h^+h^-$ decays, where $h=\pi$ or $\rho$
can be used to constrain the angle $\beta_c$ of the $cu$ unitarity triangle up to theoretical
uncertainties.  Here we discuss the theoretical uncertainty from penguin contributions
that can be mitigated through the use of isospin analyses.  We show that uncertainty from
penguin pollution on a measurement of $\beta_c$ (or alternatively the mixing phase) 
in $D^0\to \pi^+\pi^-$ ($\rho^+\rho^-$) decays is $2.7^\circ$ ($4.6^\circ$).
We also comment on the applicability of this method to $D^0\to\rho\pi$ decays for which measurements of weak phases with a precision below the one degree level may be possible.
\end{abstract}


\maketitle


}

\setcounter{footnote}{0}

\section{Introduction}
\label{sec:intro}

The Standard Model (SM) description of \CP violation is defined by a single
phase in a $3\times 3$ complex, unitary transformation known as the 
Cabibbo-Kobayashi-Maskawa
(CKM) quark-mixing matrix~\cite{Cabibbo:1963yz,Kobayashi:1973fv}.  
Measurements of time-dependent \CP asymmetries in charm decays may provide 
a means to test the SM expectations of the CKM matrix for up-type quarks, a test that has yet to be made~\cite{Bevan:2011up}.  Such measurements require knowledge of the weak mixing phase, currently only poorly determined, but one can use, for instance, the difference between measured phases in $D^0\to K^+K^-$ and $D^0\to h^+h^-$ decays
\footnote{Charge conjugation is implied throughout.}, 
where $h=\pi$ or $\rho$, to obtain a constraint on $\bteff$, related to $\beta_c$, one of the angles 
of the $cu$ unitarity triangle up to theoretical uncertainties.  This angle is, in the SM, expected to be very small ($\sim 0.035^{\circ}$) but could be enhanced by effects from new physics.  One of the sources of theoretical uncertainty that could cause $\bteff$ to differ from $\beta_c$ comes from gluonic loop (or
penguin) amplitudes contributing to the $h^+h^-$ final state with a different
weak phase from that of the dominant tree contribution.  The resulting shift, $\delta\beta_c = \beta_c-\bteff$, is referred to as ``penguin pollution".  It was noted in Ref.~\cite{Bevan:2011up} that penguin pollution in $D\to \pi\pi$ decays could be significant, by virtue of the large branching fraction of $D^0\to \pi^0\pi^0$.  One can account for the effects of 
penguins by performing an isospin ($I$-spin) analysis of the $hh$ final states 
($h=\pi$, $\rho$) in analogy with the corresponding situations found 
in \B\ decays~\cite{Aubert:2007mj,Ishino:2006if,Aubert:2007nua,Abe:2007ez}.  
Here we present a quantitative analysis of the 
penguin pollution in $D\to \pi\pi$ and $\D\to\rho\rho$ decays using existing 
experimental data, and compare these results with expectations from
the \CP self-conjugate decays $D\to \pi^+\pi^-\pi^0$,
which appear to have a small penguin contribution~\cite{CroninHennessy:2005sy,Aubert:2007ii,Gaspero:2008rs,Gaspero:2010pz,Bhattacharya:2010id}.

In general, this set of decays proceeds via tree ($T$), $W$-exchange ($E$), and three penguin amplitudes ($P_d, P_s$ and $P_b$).  The naive SM expectation is that direct \CPV\ will be very small.  In this model, the $b$-penguin phase relative to that of the tree is large ($\sim\gamma=62^{\circ}$) but its magnitude ${\cal O}(|V_{cb}V_{ub}/V_{cd}V_{ud}\alpha_s/\pi|)\sim 10^{-4}$ is tiny.  
$E$ and $P_d$ amplitudes have the same weak phase as $T$ for these modes while the $s$-penguin phase differs by only $\beta_c\simeq 0.035^0$.  
The CKM couplings for $P_d$ and $P_s$ are large but their sum is equal to the negative of that for $P_b$ ($cu$ triangle unitarity condition).  The resulting penguin amplitude is, therefore, $\sim\Vub$ with magnitude dictated by ($U$-spin) symmetry breaking $\sim(m_s^2-m_d^2)/m_c^2$, and is expected to be small. 

Prompted by evidence, earlier reported by the LHCb collaboration~\cite{Aaij:2011in} for an unexpectedly large direct \CP\ asymmetry difference ($-0.82\pm 0.24\%$) between $\Dz\to\Kp\Km$ and $\Dz\to\pip\pim$ decays, theoretical interest recently centered on the perturbative QCD properties of the light quark amplitudes $P_s$ and $P_d$ and on the possibility for enhancement by such $U$-spin symmetry breaking~\cite{Brod:2011re}.  Estimates for these effects from $\Dz\to h^+h^-$ ($h=\pi$ or $K$) decay rates~\cite{Savage:1991wt,Brod:2012ud} could be tuned to allow \CPV\ asymmetries up to a few parts per mille in these decays.  With larger available data samples of both prompt \Dz's and from \Dz's from $B$ decays, however, the LHCb estimate for this asymmetry has shrunk to $0.49\pm 0.30 (\rm stat.)\pm 0.14 (\rm syst.)$\%~\cite{Aaij:2013bra}, which is now consistent with zero.  Experimentally, measurement of the relative influence of $P$ and $T$ amplitudes in $D$ decays can be useful in understanding this confusing picture.  We will refer to ``$P$" to encompass $P_b$, $P_d$ and $P_s$ amplitudes with their QCD enhancements, weak phase $\sim\gamma$ relative to $T$ and the property of allowing only $\Delta I=1/2$ transitions (conversion of a $c$ to a $u$ quark).  In contrast, we consider ``$T$" to include tree and $W$-exchange amplitudes, which have the same weak phase and allow both $\Delta I=1/2$ and $3/2$ transitions.

Here we examine the use of methods originally developed for the corresponding \B physics problem to constrain
the loop (penguin pollution) corrections to unitarity triangle phase measurements in
$c\to d$ transitions.
Using currently available $D$ decay measurements, we then attempt to estimate the precision with which such measurements of $P$ and $T$ could be made.  

\section{Isospin analysis of $D \to\pi\pi$ decays}
\label{sec:ipipi}

\begin{figure}[!htb]
\begin{center}
  \resizebox{7.5cm}{!}{
\includegraphics{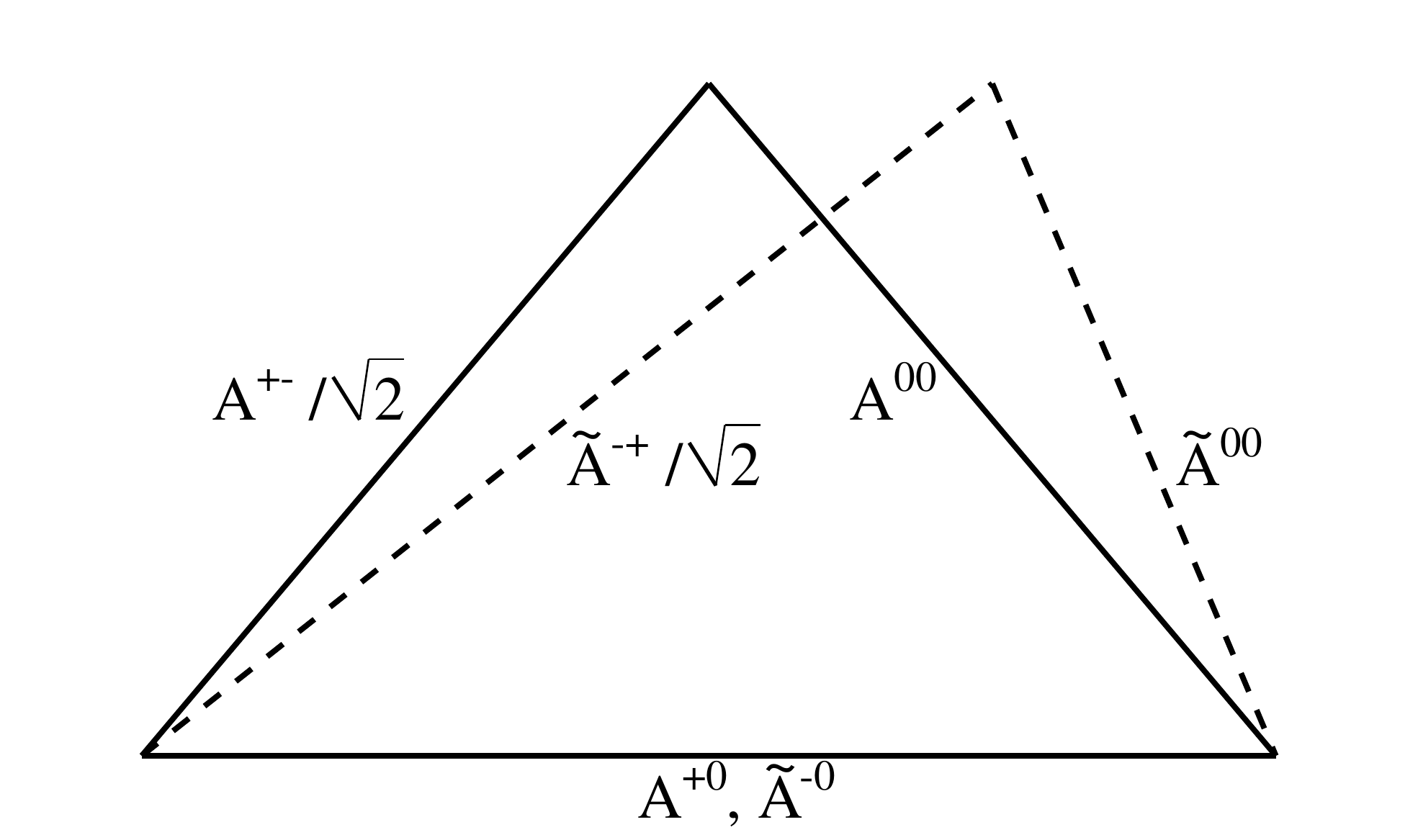}
}
  \caption{The $I$-spin triangle relations given in Eq.~(\ref{eq:isospinrelations}).}
  \label{fig:isospintriangles}
\end{center}
\end{figure}

The prescription given here parallels the one presented in Ref.~\cite{gronaulondon} which outlines how to measure the $bd$ unitarity triangle angle $\alpha$ from $\B\to \pi\pi$ decays by constraining the penguin pollution.  In this strategy, the $T$ and $P$ amplitudes are distinguished on the basis of their $I$-spin structures.  Bose symmetry dictates that, for either $B$ or $D$ decays, the two-pion (or two-$\rho$) final states can be in an $I=0$ or $I=2$ final state, calling for specific fractions of both $\Delta I = 1/2$ and $\Delta I = 3/2$ components.  $I$-spin symmetry, ignoring any small ($\sim 1\%$) electromagnetic contributions, calls for a triangular relationship between amplitudes $A^{ij}(\overline{A}^{ij})$ for $D(\Db)\to h^ih^j$ decays ($h=\pi$ or $h=\rho$): 
\begin{eqnarray}
\frac{1}{\sqrt{2}}A^{+-} = A^{+0} - A^{00},\nonumber\\
\frac{1}{\sqrt{2}}\overline{A}^{-+} = \overline{A}^{-0} - \overline{A}^{00}, \label{eq:isospinrelations}
\end{eqnarray}
where the charges are $i, j= +1, -1, 0$.  These two triangles (shown in Fig.~\ref{fig:isospintriangles}) can be aligned, by a rotation $2\beta_c$ of one of the triangles, with a common base given by $A^{+0}=\overline{A}^{-0}$ where $|A^{+0}|=|\overline{A}^{-0}|$.  By convention the rotated amplitudes are usually referred to as $\widetilde{A}^{ij}$.  Penguin operators only allow $\Delta I=1/2$ transitions so, in the limit of $I$-spin symmetry, the amplitudes $A^{+0}$ and $\overline{A}^{-0}$ are pure tree ($I=2$), 
whereas the other amplitudes are a combination of tree ($I=0,2$) and penguin ($I=0$) contributions.  
In this case the angle between $A^{+-}$ and 
$\widetilde{A}^{-+}$, the shift in the measured phase resulting from penguin contributions, 
is $2\delta\beta_c$.  Neglecting other sources of theoretical uncertainty one has $2\delta\beta_c = 2(\beta_c - \bteff)$.
A similar relation exists for time-dependent asymmetry measurements in the $h^0h^0$ mode 
between $A^{00}$ and $\widetilde{A}^{00}$.

One must measure the rates and \CP asymmetries for $D^0\to h^+h^-$, $D^\pm\to h^\pm h^0$,
and $D^0\to h^0h^0$ in order to extract the weak phase of interest, $\beta_c$.
The amplitude of sinusoidal oscillation in the time-dependence of $D\to \pi^+\pi^-$
decays is related to $\sin(\phi_{MIX}-2\bteff)$ as discussed in Ref.~\cite{Bevan:2011up},
where $\phi_{MIX}$ is the \Dz mixing phase.  The 
proposed $I$-spin analysis would enable one to translate a measurement
of $\bteff$ to a constraint on $\beta_{c}$, given a precise determination
of the mixing phase and the aforementioned $D\to hh$ amplitudes.  It would also provide an estimate of the $|P/T|$ ratio.

A noteworthy difference between $D$ and $B$ meson decays to $hh$ is the magnitude of \CP asymmetries expected.  These are much smaller, by a factor $\sim 10^{-2}$, for $D$'s so electroweak penguin processes, that violate $I$-spin conservation, could be important.  These processes do, however, preserve \CP symmetry, so by writing the relations in terms of the differences between $D$ and $\Db$ amplitudes, such effects are largely removed, at least to first order~\cite{Grossman:2012eb}.  An estimate of the uncertainties in $\delta\beta_c$ is, therefore, made ignoring EW penguins.

Current experimental constraints (i.e. branching fractions) 
on the inputs necessary to perform an $I$-spin analysis
of $D \to\pi\pi$ decays can be found in Ref.~\cite{pdg}.
%
At this time
there are no experimentally measured \CP asymmetries, however it is safe to assume in the 
SM that direct \CP asymmetries in these modes are small so that, in effect, the triangle relations of 
Eq.~(\ref{eq:isospinrelations}) are equivalent up to a possible reflection in $A^{+0}$.  
Hence there is an ambiguity in the value of $2\delta\beta_c$.  With the large samples of data in these modes
that will soon be available at BES III, \belletwo and possibly also at LHCb, one would be able to refine 
our knowledge of the inputs to this $I$-spin analysis, and eventually relax the assumptions on
\CP asymmetries.

\begin{figure}[!ht]
\begin{center}
  \resizebox{7.5cm}{!}{
\includegraphics{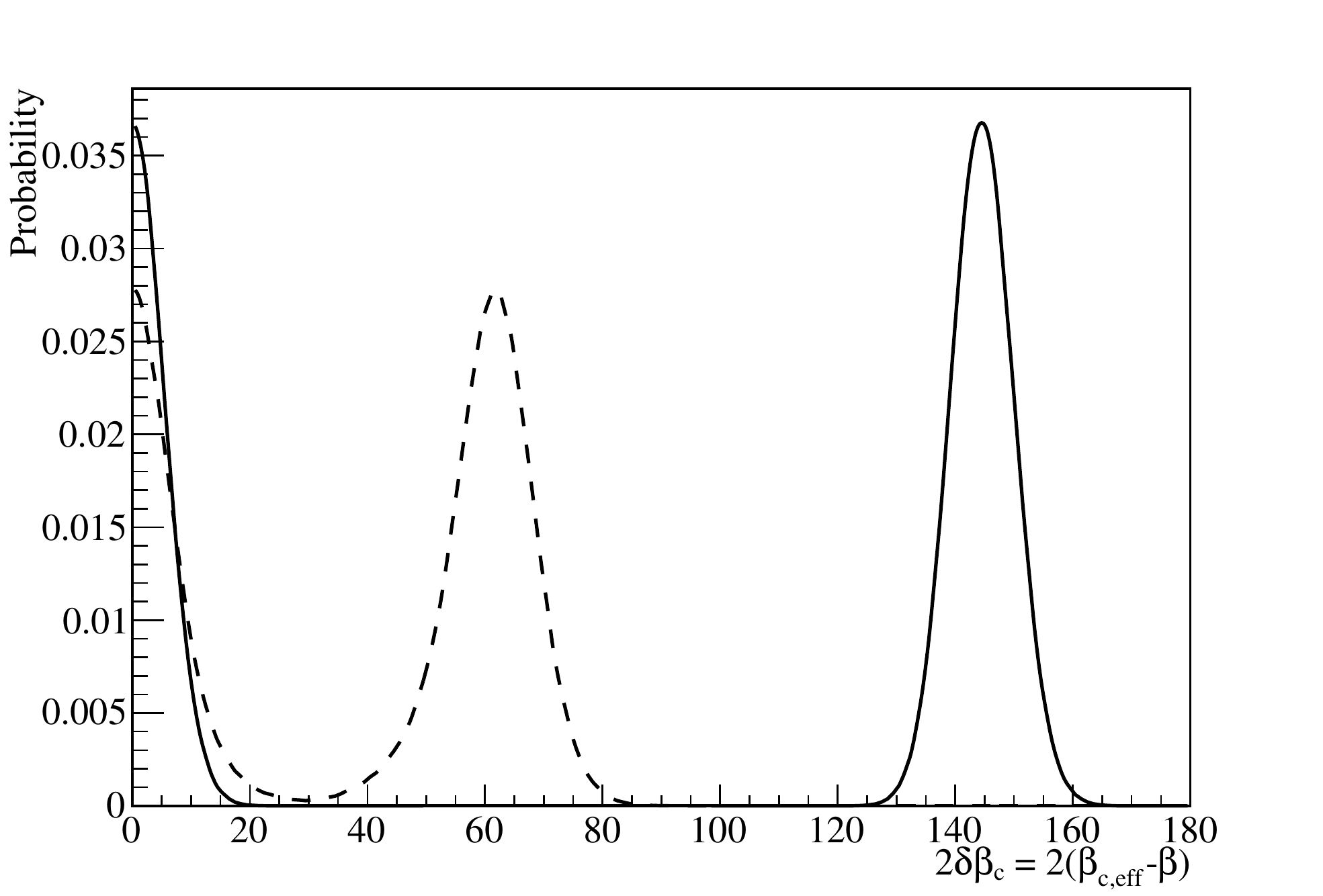}
}
  \caption{The constraint on $2\delta\beta_c$ obtained from (solid) $D\to \pi\pi$ and (dashed) $D\to \rho\rho$ 
  decays using existing data.  Mirror solutions arising from ambiguities 
  in the orientation of the $I$-spin triangles also exist for 
  $-180^\circ \leq 2\delta\beta_c < 0$.}
  \label{fig:results}
\end{center}
\end{figure}

We create ensembles of Monte Carlo simulated experiments, based on the results given in Ref.~\cite{pdg},
in order to compute $2\delta\beta_c$, noting the ambiguity in the orientation of the two relations
given in Eq.~(\ref{eq:isospinrelations}).  As a cross-check we also use the same method as in 
Ref.~\cite{Aubert:2004zra} to verify the numerical estimates obtained.
On performing the $I$-spin analysis using existing data 
one extracts the constraint shown in Fig.~\ref{fig:results}, where the 
uncertainty on each of the solutions is $5.4^\circ$.  Hence the uncertainty from penguin 
contributions on a measurement of $\beta_c$ from $D\to \pi\pi$ decays is found to be 
$2.7^\circ$ using current data. The $\piz\piz$ final state is required in the $I$-spin analysis, 
so it may not be possible to improve this estimate using data collected solely from hadron collider 
experiments.  These results rely on measurements of decay rates limited by systematic uncertainty in $\piz$ efficiency. Measurement of the \CP asymmetries of these rates are less limited, however.  
%
%
We estimate, assuming a similar detection performance is achievable to that 
found at \babar and Belle, that results from high statistics \epem based experiments 
could be able to measure $\delta\beta_c$ at the level of ${\cal O}(1.3^{\circ})$
in $D\to \pi\pi$ decays.
Possible improvements in tracking and calorimetry should also be investigated to
understand if a sub-$1^\circ$ level could be achievable.

\section{Isospin analysis of $D \to\rho\rho$ decays}
\label{sec:irhorho}

The $I$-spin analysis for $D \to\rho\rho$ decays parallels the procedure described
for $D \to\pi\pi$, with the exception that there may be an $I=1$ contribution
to the (4-pion) final states as discussed in Ref.~\cite{falk} for the corresponding $B \to\rho\rho$ case.
It is possible to test for an $I=1$ component by measuring 
$\bteff$ as a function of the $\pi\pi$ invariant mass in the $\rho$ region.  Any observed
variation as a function of the difference in these masses would be an indication of such a contribution.

Experimentally the situation is not quite as straight forward as the $D\to\pi\pi$
case.  The
presence of two broad $\rho$ resonances in the final state will result in larger backgrounds
than in the $\pi\pi$ case.  Furthermore the four pion $\rho\rho$ final state has to be 
distinguished from other possible resonant and non-resonant contributions, taking into 
account any interference that may occur. In particular there could be visible interference
effects between $\rho^0$ and $\omega$ apparent in the $\rho^0$ line shape for the $\rho^+\rhoz$ and
$\rhoz\rhoz$ modes. As noted in Ref.~\cite{Bevan:2011up}, we expect
the fraction of longitudinally polarized events \fl to be non-trivial ($\fl \sim 0.83$), so one has to 
perform a transversity analysis in order to extract \CP even and \CP odd components 
of the final state. 
We note that experimental constraints on \fl for $\rho^0\rho^0$
are consistent with naive expectations discussed in Ref.~\cite{Bevan:2011up}.
Having done this, one can then perform an $I$-spin analysis
for each of the three signal components (i.e. the three transversity amplitudes), where in principle
one could have a different level of penguin pollution for each component.
Current experimental constraints on the inputs necessary to perform an $I$-spin analysis
of $D \to\rho\rho$ decays can be found in Ref.~\cite{pdg}, 
where we 
interpret the four pion final state as being dominated by $\rho\rho$ decays.  As before
it is safe to assume that direct \CP asymmetries in the $\rho\rho$ modes are small.
In the following we assume that only the longitudinally polarized events are used in order
to constrain $\bteff$.  
On performing the $I$-spin analysis one extracts the constraint on $2\delta\beta_c$ shown as the dashed
line in Fig.~\ref{fig:results}.  The penguin uncertainty is
broader than obtained for the $\pi\pi$ case, and for $\rho\rho$ the two different solutions 
have an uncertainty of $9.2^\circ$ and overlap slightly.  Hence there is an uncertainty of
$4.6^{\circ}$ on a future measurement of $\beta_c$ from this source.  
The fraction of resonant $\rho\rho$ events required to minimize the level of unphysical
results is one, highlighting the need for improved experimental input in this area.
As with the $\pi\pi$ case one needs to reconstruct
final states with neutral particles in order to perform this $I$-spin analysis.
Thus it is important that experiments such as BES III and \belletwo study 
the $\D\to \rho\rho$ final states in detail in order to provide the appropriate inputs
for the $I$-spin analysis.   However, we expect that LHCb should be able to
contribute significantly via measurement of $\Dz\to\rho^0\rho^0$ decays.



\section{Isospin analysis of $D \to \pi^+\pi^-\pi^0$ decays}
\label{sec:i3pi}

Time-dependent \CP asymmetry measurements of \Dz decays to the \CP-even state $\rho^0\piz$ also provide information on $\bteff$~\cite{Bevan:2011up}.  These decays have contributions from Cabibbo suppressed exchange ($C$) and color suppressed tree diagrams ($T$), each with the weak phase of $\vcd^*\vud$ 
(following the usual convention) that could contribute to a \CP asymmetry.  The penguin diagram 
can again play a role in the weak phase and some estimate of it needs to be made.  

The $\rho^0\piz$ mode represents almost 30\% of the 3-body decays $\Dz\to\pi^+\pi^-\pi^0$ whose branching fraction is $(1.43\pm 0.06)\%$, over ten times larger than that for $\Dz\to \pi^+\pi^-$.  The advantage arising from the additional data, despite an increase in experimental complexity of the $I$-spin makeup of the $\rho\pi$ states, is actually further enhanced by additional information from the other $\rho\pi$ charge modes and can be used to place interesting constraints on the value for $\beta_c$ directly.  

Amplitude analyses have been made by the CLEO-c~\cite{CroninHennessy:2005sy} and \babar~\cite{Aubert:2007ii} collaborations.  The \babar analysis, which used a sample almost 20 times larger than CLEO's, revealed, in addition to the 3 $\rho\pi$ modes, the presence of radial excitations of the $\rho$ as well as other \CP-odd eigenstates ($f_{\sst 0}\piz$ and $f_{\sst 2}\piz$ with $I=1)$.  An $I$-spin study~\cite{Gaspero:2008rs,Gaspero:2010pz} of the \babar amplitudes has also shown that $I=0$ is dominant in the three $\Dz\to\rho\pi$ modes.  This somewhat surprising observation has been interpreted as being consistent with $\su3f$ relationships~\cite{Bhattacharya:2010id} with only $C$ and $T$ amplitudes from other $\Dz\to P+V$ (pseudoscalar~+~vector) decays.  The effect of penguin contributions was not required in this interpretation, however.  Nevertheless, larger event samples are yet to come from all these channels that may be able to clarify the role of penguins further.

\begin{figure}[!ht]
\begin{center}
  \resizebox{7.5cm}{!}{
\includegraphics{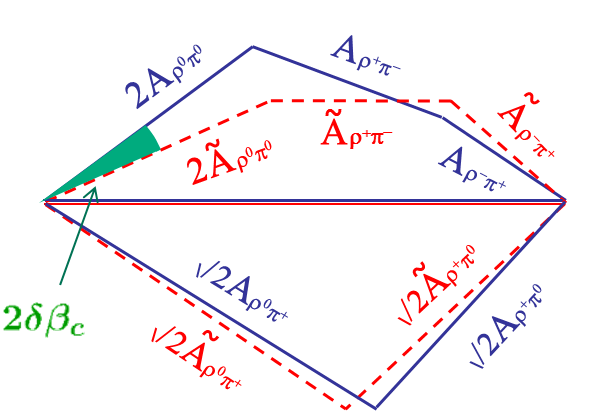}
}
\caption{The $I$-spin pentagon relation for the $D\to\rho\pi$ modes given in Eq.~(\ref{eq:pentagon}).  Amplitudes for \Dz\ (solid) and \Dzb\ (dotted) are shown.  The \Dzb\ pentagon is rotated so that the sums on each side of Eq.~(\ref{eq:pentagon}) coincide with those for \Dz decays.  These rotated amplitudes are labeled with a tilde.  The effect of penguins that can contribute a change, $\delta\beta_c$, in the phase of the $\rho^0\piz$ amplitude is illustrated.}
\label{fig:pentagon}
\end{center}
\end{figure}

The $I$-spin structure of the $\rho\pi$ final states accessible to \Dz or $D^{\pm}$ decays differs from that of either $\pi\pi$ or $\rho\rho$ states~\cite{Gronau:2004sj}, each with pairs of identical Bosons allowing predominantly even $I$ values.  States with $I=0, 1$ or $2$ are now allowed, and all but $I=2$ are accessible to penguin transitions.  The decay amplitudes can be written~\cite{Grossman:2012eb} in terms of $I$-spin amplitudes ${\cal A}_1$ and ${\cal B}_1$, $\Delta I=1/2$ transitions, respectively, to $I=1$ and $I=0$ three pion states, and ${\cal A}_3$ and ${\cal B}_3$, $(\Delta I=3/2)$ transitions to $I=2$ and $I=1$ states.  SM penguins cannot contribute to $\Delta I=3/2$ processes and we ignore $\Delta I=5/2$ transitions that are not expected in the SM.  Then
\bea
  A^{+-}  &=& {\cal A}_3 + {\cal B}_3 + \frac{1}{\sqrt 2}{\cal A}_1 + {\cal B}_1\\
                &=& T^{+-}+P_1+P_0 \\
  A^{-+}  &=& {\cal A}_3 - {\cal B}_3 - \frac{1}{\sqrt 2}{\cal A}_1 + {\cal B}_1\\
                &=& T^{-+}-P_1+P_0 \\
  A^{00 } &=& 2{\cal A}_3 - {\cal B}_1\\
                &=& [T^{+-}+T^{-+}-T^{+0}-T^{0+}]/2-P_0 \\
  A^{+ 0} &=& \frac{3}{\sqrt 2}{\cal A}_3 - \frac{1}{\sqrt 2}{\cal B}_1 + {\cal A}_1\\
                &=& [T^{+0}+2P_1 ]/\sqrt 2\\
  A^{0 +} &=& \frac{3}{\sqrt 2}{\cal A}_3 + \frac{1}{\sqrt 2}{\cal B}_1 - {\cal A}_1\\
                &=& [T^{0+}-2P_1 ]/\sqrt 2
\eea
where the first superscript in the decay $A$ and tree $T$ amplitudes is the charge state for the $\rho$, and the second is that for the $\pi$.  $P_0$ and $P_1$ are, respectively, penguin amplitudes leading to $I=0$ and $I=1$ states.  A sum rule involving all five decay amplitudes
\begin{eqnarray}
\label{eq:pentagon}
  A^{+-} + A^{-+} + 2A^{00} &=& 
  \sqrt 2 (A^{+0} + A^{0+})
\end{eqnarray}
can be inferred, and is represented graphically in Fig.~\ref{fig:pentagon}.  The corresponding antiparticle amplitudes $\widetilde{A}$ are similarly related though they each can differ from the $\Dz$ ones.  Each side of Eq.~(\ref{eq:pentagon}) corresponds to a sum of amplitudes with $\Delta I=3/2$ (tree only).  The asymmetry between these sums for $\Dz$ and $\Dzb$ should therefore be negligible in the SM.  Even if $I$-spin symmetry is broken, the difference between the $\Dz$ and $\Dzb$ sums should be very small~\cite{Grossman:2012eb}.  These sums, therefore, play a similar role to the $\pip\piz$ and $\rho^+\rho^0$ modes discussed earlier, and a similar rotation of the charged $\overline D$ amplitudes through $2\bteff$, illustrated in Fig.~\ref{fig:pentagon}, aligns them.  Penguin contributions lead to a difference in phase $2\delta\beta_c$ to $\widetilde{A}^{00}$ (the rotated $\Dzb\to\rho^0\piz$ amplitude) relative to $A^{00}$.

%
%

Charged $D\to\rho\pi$ decays are observed in two different Dalitz plots, $\Dp\to\pip\piz\piz$ ($\rho^+\piz$) and $\Dp\to\pip\pim\piz$ ($\rho^0\pip$).  It is, therefore, impossible to determine their phases relative to each other, or to any of the three neutral $\rho\pi$ modes.  For a similar reason, it is difficult to relate their magnitudes in light of uncertainties in the normalization of amplitudes from these Dalitz plots.  The charged modes, therefore, can only add a somewhat weak $I$-spin constraint.  Fortunately, as pointed out by Quinn and Snyder~\cite{Snyder:1993mx}, a time-dependent amplitude analysis of the $\pip\pim\piz$ Dalitz plot alone (just the upper part of Fig.~\ref{fig:pentagon}) is sufficient to extract the weak decay phase $\bteff$, as well as the $P/T$ ratio from the time-dependence of all three $\Dz\to\rho\pi$ modes.  The relative phase of the amplitudes $A^{00}$ and $\widetilde{A}^{00}$ is $2\delta\beta_c$, $\bteff$ can also be determined, and is obtained from the time-dependence of the $\Dz(\Dzb)\to\rho^0\piz$ amplitude and $\delta\beta_c$ from the $\rho^{\pm}$ modes.  

The phase $\beta_c$ can also be obtained directly, with the same precision as that in $\delta\beta_c$.  A further advantage of analyzing just the three $B^0\to(\rho\pi)^0$ channels is that only their amplitude ratios and relative phases are required, and information on these reside entirely within the same Dalitz plot.  The precision of their measurements, therefore, should all be reduced with larger sample sizes, with no limiting systematic uncertainty from $\piz$ detection efficiency.

To estimate the precision in $2\delta\beta_c$, we take results from the \babar time-integrated amplitude analysis of $\Dz\to\pip\pim\piz$ decays~\cite{Aubert:2007ii}.  At this statistical level, interferences in the Dalitz plot allow measurement precision of relative phases of $A^{-+}$, $A^{+-}$ and $A^{00}$ of $\sim 1^{\circ}$ and of relative magnitudes at the level of $\sim 1\%$.  
The \babar analysis combined both $\Dz$ and $\Dzb$ samples in a single fit so, in our study, we make the assumption that central values for both sets of amplitudes are the same (i.e. no direct \CPV).  Our expectation is that $\delta\beta_c=0$.  In a time-dependent analysis, mixing brings the $\Dz$ and $\Dzb$ into interference, allowing measurement of their relative decay phases.  The time-integrated \babar analysis did not, therefore, provide any information from mixing-induced \CPV on the central value for $\bteff$, so we set this to be zero.

\begin{figure}[!ht]
\begin{center}
 \resizebox{7.5cm}{!}{\includegraphics{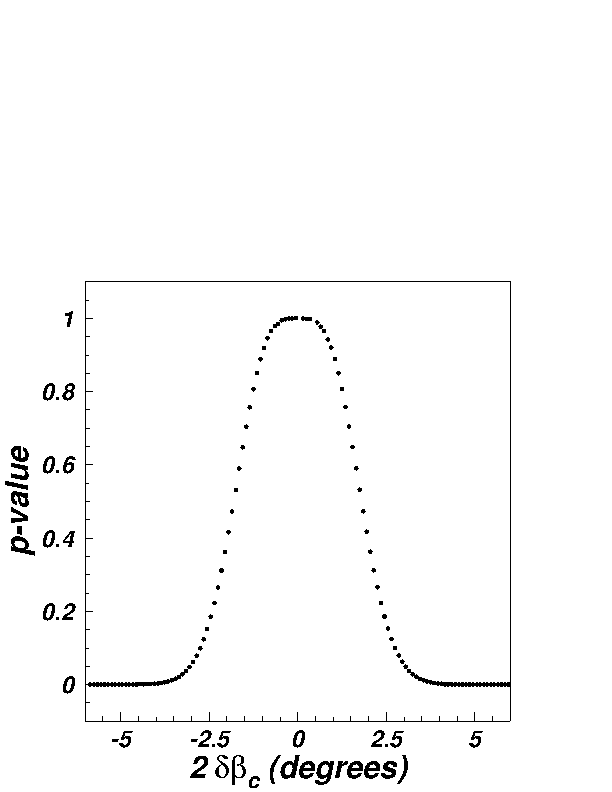}}
 \caption{$p$-values for minimum $\chi^2$ fits to \babar data for $2\delta\beta_c$ about zero, the value expected as   outlined in the text.  A second, identical peak centered at $180^{\circ}$ is also observed, but not shown here.}
 \label{fig:rhopiscan}
\end{center}
\end{figure}

Following the procedure used by the $B$ factories (for example in \cite{Lees:2013nwa}) we extract from the \babar data, $P$ and $T$ amplitudes with strong and weak phases for a set of values for $2\delta\beta_c$ around $0^{\circ}$.  At each value, we obtain the p-value for the minimum $\chi^2$ for these parameters.  These p-values are plotted in Fig.~\ref{fig:rhopiscan}.  As expected, a peak centered at $0^{\circ}$ is seen with a width of close to $1^{\circ}$.  
%


So far, we have implicitly assumed that the \Dz mixing
parameters are well known, as we anticipate they will be when
the analyses outlined here are made.  Perhaps the mixing phase
($\phi_{MIX}=arg[q/p]$) is one of these parameters that most
directly affects any conclusions on possible constraints on
the $cu$ unitarity triangle.  This phase is not well understood 
theoretically, and the time-dependent \CP asymmetry measurements
discussed here determine the combination $\phi_{MIX}-2\beta_c$.
It is estimated~\cite{O'Leary:2010af} that $\phi_{MIX}$ can be
measured in a Super Flavor Factory with precision of order $1^\circ$.
One could therefore conclude that, with planned facilities,
measurements of the $cu$ unitarity triangle will be not be
possible, but deviations from the SM expectations can be
constrained to within a few degrees.

\section{Summary}
\label{sec:summary}

The analysis of penguin pollution discussed here is the first step toward resolving the theoretical uncertainties
associated with a measurement of the mixing phase in the near term, and ultimately the $cu$ unitarity 
triangle phase $\beta_c$.  Additional theoretical 
uncertainties should be investigated, including those related to long distance amplitudes contributing to the 
direct decay of a \Dz or \Dzb meson into the final state, as well as possible $I$-spin breaking terms
arising from electroweak penguin contributions.
As noted in Ref.~\cite{Bevan:2011up}, the penguin contribution in $D\to\pi\pi$ decays is potentially large,
however current experimental inputs are precise enough to enable one to constrain the uncertainty from 
penguins on $\beta_c$ to a few degrees.  \sffs should be able to further reduce this uncertainty 
with improved measurements of the $\pi\pi$ final states.
Assuming that the four-pion final states measured in charged and neutral decays of $D$ mesons are
dominated by resonant $\rho\rho$ contributions, we find that the maximum value of the 
penguin pollution contributions in 
$\D\to \rho\rho$ decays is smaller than that for $\pi\pi$, however current experimental data are of
limited value and we obtain a worse precision for the penguin pollution uncertainty.  A detailed
experimental analysis of all of the $\rho\rho$ final states is needed in order to improve on the estimate
we obtain here.
While BES III will not be able to contribute to the time-dependent asymmetry inputs required to extract
an estimate of $\beta_c$, this experiment should be able to make significant contributions to the knowledge
of time integrated rates and asymmetries used as inputs to the $I$-spin analysis.
Time-dependent analyses are likely to provide measurements of $\bteff$ from $\Dz\to\rho^0\piz$ decays with precision in the $1^0$ range using the large samples expected from the Super Flavor Factories.  In the meantime,
BES III should be able to provide large samples of such decays allowing for more precise studies to be made of time-integrated Dalitz plots.  It is also possible that \CP asymmetries and $I$-spin analyses of the various charge states of the $\rho\pi$ systems can be used to estimate penguin contributions.  

The $\Dz\to\rho\pi$ modes are especially encouraging.  In particular, full time-dependent amplitude analyses of the neutral $\Dz\to 3\pi$ Dalitz plots should allow internally compatible measurements for magnitudes and phases for each of the three charge modes for both $\Dz$ and $\Dzb$ decays.  These measurements should not be limited by systematic uncertainties in $\piz$ detection efficiency, so resolution in $\delta\beta_c$ could shrink considerably below the $1^{\circ}$ level as larger samples are used.

The different constraints on penguin pollution evident in Figs~\ref{fig:results} and \ref{fig:rhopiscan} 
are a clear indication that some of the ambiguities involved in a measurement of 
$\beta_c$ from $D\to hh$ decays can be resolved
when results from different final states are combined.  A recent result from Lattice QCD
by the RBC-UKQCD collaboration~\cite{Boyle:2012ys} states that penguins should be small in this
system.  This prediction can be tested using time-dependent asymmetry measurements of $D\to hh$
decays combined with the isospin analysis discussed here. We conclude that with sufficient
data it is feasible to perform a measurement of the $cu$ unitarity triangle phase $\beta_{c}$ to
a precision comparable to that achievable by the \B Factories and LHCb 
for the $bd$ ($bs$) unitarity triangle phase $\beta$ ($\phi_s$).

\section{Acknowledgments}

This work has been supported by the US National Science Foundation,
under grant number PHY-0757876.  BM also wishes to thank the University of Oxford and the Leverhulme Foundation for their generous support as a visiting Professor in Oxford where much of this work was done.  We have adapted the numerical algorithm originally developed by C.~Yeche for Ref.~\cite{Aubert:2004zra} to cross-check the numerical estimates summarized here.

\bibliography{note}

\begin{thebibliography}{27}
\expandafter\ifx\csname natexlab\endcsname\relax\def\natexlab#1{#1}\fi
\expandafter\ifx\csname bibnamefont\endcsname\relax
  \def\bibnamefont#1{#1}\fi
\expandafter\ifx\csname bibfnamefont\endcsname\relax
  \def\bibfnamefont#1{#1}\fi
\expandafter\ifx\csname citenamefont\endcsname\relax
  \def\citenamefont#1{#1}\fi
\expandafter\ifx\csname url\endcsname\relax
  \def\url#1{\texttt{#1}}\fi
\expandafter\ifx\csname urlprefix\endcsname\relax\def\urlprefix{URL }\fi
\providecommand{\bibinfo}[2]{#2}
\providecommand{\eprint}[2][]{\url{#2}}

\bibitem[{\citenamefont{Cabibbo}(1963)}]{Cabibbo:1963yz}
\bibinfo{author}{\bibfnamefont{N.}~\bibnamefont{Cabibbo}},
  \bibinfo{journal}{Phys. Rev. Lett.} \textbf{\bibinfo{volume}{10}},
  \bibinfo{pages}{531} (\bibinfo{year}{1963}).

\bibitem[{\citenamefont{Kobayashi and Maskawa}(1973)}]{Kobayashi:1973fv}
\bibinfo{author}{\bibfnamefont{M.}~\bibnamefont{Kobayashi}} \bibnamefont{and}
  \bibinfo{author}{\bibfnamefont{T.}~\bibnamefont{Maskawa}},
  \bibinfo{journal}{Prog. Theor. Phys.} \textbf{\bibinfo{volume}{49}},
  \bibinfo{pages}{652} (\bibinfo{year}{1973}).

\bibitem[{\citenamefont{Bevan et~al.}(2011)\citenamefont{Bevan, Inguglia, and
  Meadows}}]{Bevan:2011up}
\bibinfo{author}{\bibfnamefont{A.}~\bibnamefont{Bevan}},
  \bibinfo{author}{\bibfnamefont{G.}~\bibnamefont{Inguglia}}, \bibnamefont{and}
  \bibinfo{author}{\bibfnamefont{B.}~\bibnamefont{Meadows}},
  \bibinfo{journal}{Phys. Rev.} \textbf{\bibinfo{volume}{D84}},
  \bibinfo{pages}{114009} (\bibinfo{year}{2011}), \eprint{1106.5075}.

\bibitem[{\citenamefont{Aubert et~al.}(2007{\natexlab{a}})}]{Aubert:2007mj}
\bibinfo{author}{\bibfnamefont{B.}~\bibnamefont{Aubert}} \bibnamefont{et~al.}
  (\bibinfo{collaboration}{\babar}), \bibinfo{journal}{Phys. Rev. Lett.}
  \textbf{\bibinfo{volume}{99}}, \bibinfo{pages}{021603}
  (\bibinfo{year}{2007}{\natexlab{a}}), \eprint{hep-ex/0703016}.

\bibitem[{\citenamefont{Ishino et~al.}(2007)}]{Ishino:2006if}
\bibinfo{author}{\bibfnamefont{H.}~\bibnamefont{Ishino}} \bibnamefont{et~al.}
  (\bibinfo{collaboration}{Belle}), \bibinfo{journal}{Phys. Rev. Lett.}
  \textbf{\bibinfo{volume}{98}}, \bibinfo{pages}{211801}
  (\bibinfo{year}{2007}), \eprint{hep-ex/0608035}.

\bibitem[{\citenamefont{Aubert et~al.}(2007{\natexlab{b}})}]{Aubert:2007nua}
\bibinfo{author}{\bibfnamefont{B.}~\bibnamefont{Aubert}} \bibnamefont{et~al.}
  (\bibinfo{collaboration}{Babar}), \bibinfo{journal}{Phys. Rev.}
  \textbf{\bibinfo{volume}{D76}}, \bibinfo{pages}{052007}
  (\bibinfo{year}{2007}{\natexlab{b}}), \eprint{0705.2157}.

\bibitem[{\citenamefont{Somov et~al.}(2007)}]{Abe:2007ez}
\bibinfo{author}{\bibfnamefont{A.}~\bibnamefont{Somov}} \bibnamefont{et~al.}
  (\bibinfo{collaboration}{Belle}), \bibinfo{journal}{Phys. Rev.}
  \textbf{\bibinfo{volume}{D76}}, \bibinfo{pages}{011104}
  (\bibinfo{year}{2007}), \eprint{hep-ex/0702009}.

\bibitem[{\citenamefont{Cronin-Hennessy et~al.}(2005)}]{CroninHennessy:2005sy}
\bibinfo{author}{\bibfnamefont{D.}~\bibnamefont{Cronin-Hennessy}}
  \bibnamefont{et~al.} (\bibinfo{collaboration}{CLEO Collaboration}),
  \bibinfo{journal}{Phys. Rev.} \textbf{\bibinfo{volume}{D72}},
  \bibinfo{pages}{031102} (\bibinfo{year}{2005}), \eprint{hep-ex/0503052}.

\bibitem[{\citenamefont{Aubert et~al.}(2007{\natexlab{c}})}]{Aubert:2007ii}
\bibinfo{author}{\bibfnamefont{B.}~\bibnamefont{Aubert}} \bibnamefont{et~al.}
  (\bibinfo{collaboration}{BaBar Collaboration}), \bibinfo{journal}{Phys. Rev.
  Lett.} \textbf{\bibinfo{volume}{99}}, \bibinfo{pages}{251801}
  (\bibinfo{year}{2007}{\natexlab{c}}), \eprint{hep-ex/0703037}.

\bibitem[{\citenamefont{Gaspero et~al.}(2008)\citenamefont{Gaspero, Meadows,
  Mishra, and Soffer}}]{Gaspero:2008rs}
\bibinfo{author}{\bibfnamefont{M.}~\bibnamefont{Gaspero}},
  \bibinfo{author}{\bibfnamefont{B.}~\bibnamefont{Meadows}},
  \bibinfo{author}{\bibfnamefont{K.}~\bibnamefont{Mishra}}, \bibnamefont{and}
  \bibinfo{author}{\bibfnamefont{A.}~\bibnamefont{Soffer}},
  \bibinfo{journal}{Phys. Rev.} \textbf{\bibinfo{volume}{D78}},
  \bibinfo{pages}{014015} (\bibinfo{year}{2008}), \eprint{0805.4050}.

\bibitem[{\citenamefont{Gaspero}(2010)}]{Gaspero:2010pz}
\bibinfo{author}{\bibfnamefont{M.}~\bibnamefont{Gaspero}},
  \bibinfo{journal}{AIP Conf.Proc.} \textbf{\bibinfo{volume}{1257}},
  \bibinfo{pages}{242} (\bibinfo{year}{2010}), \eprint{1001.3317}.

\bibitem[{\citenamefont{Bhattacharya et~al.}(2010)\citenamefont{Bhattacharya,
  Chiang, and Rosner}}]{Bhattacharya:2010id}
\bibinfo{author}{\bibfnamefont{B.}~\bibnamefont{Bhattacharya}},
  \bibinfo{author}{\bibfnamefont{C.-W.} \bibnamefont{Chiang}},
  \bibnamefont{and} \bibinfo{author}{\bibfnamefont{J.~L.}
  \bibnamefont{Rosner}}, \bibinfo{journal}{Phys. Rev.}
  \textbf{\bibinfo{volume}{D81}}, \bibinfo{pages}{096008}
  (\bibinfo{year}{2010}), \eprint{1004.3225}.

\bibitem[{\citenamefont{Aaij et~al.}(2012)}]{Aaij:2011in}
\bibinfo{author}{\bibfnamefont{R.}~\bibnamefont{Aaij}} \bibnamefont{et~al.}
  (\bibinfo{collaboration}{LHCb Collaboration}), \bibinfo{journal}{Phys. Rev.
  Lett.} \textbf{\bibinfo{volume}{108}}, \bibinfo{pages}{111602}
  (\bibinfo{year}{2012}), \eprint{1112.0938}.

\bibitem[{\citenamefont{Brod et~al.}(2012{\natexlab{a}})\citenamefont{Brod,
  Kagan, and Zupan}}]{Brod:2011re}
\bibinfo{author}{\bibfnamefont{J.}~\bibnamefont{Brod}},
  \bibinfo{author}{\bibfnamefont{A.~L.} \bibnamefont{Kagan}}, \bibnamefont{and}
  \bibinfo{author}{\bibfnamefont{J.}~\bibnamefont{Zupan}},
  \bibinfo{journal}{Phys.Rev.} \textbf{\bibinfo{volume}{D86}},
  \bibinfo{pages}{014023} (\bibinfo{year}{2012}{\natexlab{a}}),
  \eprint{1111.5000}.

\bibitem[{\citenamefont{Savage}(1991)}]{Savage:1991wt}
\bibinfo{author}{\bibfnamefont{M.~J.} \bibnamefont{Savage}},
  \bibinfo{journal}{Phys.Lett.} \textbf{\bibinfo{volume}{B259}},
  \bibinfo{pages}{135} (\bibinfo{year}{1991}).

\bibitem[{\citenamefont{Brod et~al.}(2012{\natexlab{b}})\citenamefont{Brod,
  Grossman, Kagan, and Zupan}}]{Brod:2012ud}
\bibinfo{author}{\bibfnamefont{J.}~\bibnamefont{Brod}},
  \bibinfo{author}{\bibfnamefont{Y.}~\bibnamefont{Grossman}},
  \bibinfo{author}{\bibfnamefont{A.~L.} \bibnamefont{Kagan}}, \bibnamefont{and}
  \bibinfo{author}{\bibfnamefont{J.}~\bibnamefont{Zupan}}
  (\bibinfo{year}{2012}{\natexlab{b}}), \eprint{1203.6659}.

\bibitem[{\citenamefont{Aaij et~al.}(2013)}]{Aaij:2013bra}
\bibinfo{author}{\bibfnamefont{R.}~\bibnamefont{Aaij}} \bibnamefont{et~al.}
  (\bibinfo{collaboration}{LHCb collaboration}) (\bibinfo{year}{2013}),
  \eprint{1303.2614}.

\bibitem[{\citenamefont{Gronau and London}(1990)}]{gronaulondon}
\bibinfo{author}{\bibfnamefont{M.}~\bibnamefont{Gronau}} \bibnamefont{and}
  \bibinfo{author}{\bibfnamefont{D.}~\bibnamefont{London}},
  \bibinfo{journal}{Phys. Rev. Lett.} \textbf{\bibinfo{volume}{65}},
  \bibinfo{pages}{3381} (\bibinfo{year}{1990}).

\bibitem[{\citenamefont{Grossman et~al.}(2012)\citenamefont{Grossman, Kagan,
  and Zupan}}]{Grossman:2012eb}
\bibinfo{author}{\bibfnamefont{Y.}~\bibnamefont{Grossman}},
  \bibinfo{author}{\bibfnamefont{A.~L.} \bibnamefont{Kagan}}, \bibnamefont{and}
  \bibinfo{author}{\bibfnamefont{J.}~\bibnamefont{Zupan}}
  (\bibinfo{year}{2012}), \eprint{1204.3557}.

\bibitem[{\citenamefont{Beringer et~al.}(2012)}]{pdg}
\bibinfo{author}{\bibfnamefont{J.}~\bibnamefont{Beringer}} \bibnamefont{et~al.}
  (\bibinfo{collaboration}{Particle Data Group}), \bibinfo{journal}{Phys. Rev.
  D} \textbf{\bibinfo{volume}{86}}, \bibinfo{pages}{010001}
  (\bibinfo{year}{2012}),
  \urlprefix\url{http://link.aps.org/doi/10.1103/PhysRevD.86.010001}.

\bibitem[{\citenamefont{Aubert et~al.}(2004)}]{Aubert:2004zra}
\bibinfo{author}{\bibfnamefont{B.}~\bibnamefont{Aubert}} \bibnamefont{et~al.}
  (\bibinfo{collaboration}{\babar}), \bibinfo{journal}{Phys. Rev. Lett.}
  \textbf{\bibinfo{volume}{93}}, \bibinfo{pages}{231801}
  (\bibinfo{year}{2004}), \eprint{hep-ex/0404029}.

\bibitem[{\citenamefont{Falk et~al.}(2004)}]{falk}
\bibinfo{author}{\bibfnamefont{A.~F.} \bibnamefont{Falk}} \bibnamefont{et~al.},
  \bibinfo{journal}{Phys. Rev. D} \textbf{\bibinfo{volume}{69}},
  \bibinfo{pages}{011502} (\bibinfo{year}{2004}).

\bibitem[{\citenamefont{Gronau et~al.}(2005)\citenamefont{Gronau, Lunghi, and
  Wyler}}]{Gronau:2004sj}
\bibinfo{author}{\bibfnamefont{M.}~\bibnamefont{Gronau}},
  \bibinfo{author}{\bibfnamefont{E.}~\bibnamefont{Lunghi}}, \bibnamefont{and}
  \bibinfo{author}{\bibfnamefont{D.}~\bibnamefont{Wyler}},
  \bibinfo{journal}{Phys. Lett.} \textbf{\bibinfo{volume}{B606}},
  \bibinfo{pages}{95} (\bibinfo{year}{2005}), \eprint{hep-ph/0410170}.

\bibitem[{\citenamefont{Snyder and Quinn}(1993)}]{Snyder:1993mx}
\bibinfo{author}{\bibfnamefont{A.~E.} \bibnamefont{Snyder}} \bibnamefont{and}
  \bibinfo{author}{\bibfnamefont{H.~R.} \bibnamefont{Quinn}},
  \bibinfo{journal}{Phys. Rev.} \textbf{\bibinfo{volume}{D48}},
  \bibinfo{pages}{2139} (\bibinfo{year}{1993}).

\bibitem[{\citenamefont{Lees et~al.}(2013)}]{Lees:2013nwa}
\bibinfo{author}{\bibfnamefont{J.}~\bibnamefont{Lees}} \bibnamefont{et~al.}
  (\bibinfo{collaboration}{BaBar Collaboration}) (\bibinfo{year}{2013}),
  \eprint{1304.3503}.

\bibitem[{\citenamefont{O'Leary et~al.}(2010)}]{O'Leary:2010af}
\bibinfo{author}{\bibfnamefont{B.}~\bibnamefont{O'Leary}} \bibnamefont{et~al.}
  (\bibinfo{collaboration}{SuperB}) (\bibinfo{year}{2010}), \eprint{1008.1541}.

\bibitem[{\citenamefont{Boyle et~al.}(2013)}]{Boyle:2012ys}
\bibinfo{author}{\bibfnamefont{P.}~\bibnamefont{Boyle}} \bibnamefont{et~al.}
  (\bibinfo{collaboration}{RBC Collaboration, UKQCD Collaboration}),
  \bibinfo{journal}{Phys.Rev.Lett.} \textbf{\bibinfo{volume}{110}},
  \bibinfo{pages}{152001} (\bibinfo{year}{2013}), \eprint{1212.1474}.

\end{thebibliography}



\end{document}